\documentclass[12pt,final]{iopart}
\usepackage{graphics}
\newcommand{\bea}{\begin{eqnarray}}
\newcommand{\eea}{\end{eqnarray}}
\newcommand{\be}{\begin{equation}}
\newcommand{\ee}{\end{equation}}

\def\be{\begin{eqnarray}}
\def\ee{\end{eqnarray}}
\def\bd{\begin{displaymath}}
\def\ed{\end{displaymath}}

\begin{document}
\title[RMF study of neutron rich C and Be isotopes]
{Relativistic mean field study of neutron rich even-even C and Be isotopes}
\author{G. Gangopadhyay\dag 
\footnote[3]{Corresponding author (gautam@cucc.ernet.in)}
and Subinit Roy\ddag}

\address{\dag Department of Physics, University of Calcutta,
92 Acharya Prafulla Chandra Road, Kolkata-700 009, India\\
\ddag Saha Institute of Nuclear Physics, Block AF, Sector 1, 
Kolkata- 700 064, India}

\begin{abstract}
Ground state properties of neutron rich even-even Be and C nuclei have been 
investigated using Relativistic Mean Field 
approach in co-ordinate space. The positions of the neutron drip line are 
correctly predicted for both the elements. The nucleus $^{14}$Be shows 
a two neutron halo but, contrary to expectation, $^{22}$C does not exhibit
any halo structure. In carbon nuclei, N=16 comes out as a new 
magic number. The single particle 
level ordering observed in stable
nuclei is found to be modified in neutron rich Be isotopes. 
Elastic partial scattering cross sections for proton scattering in inverse 
kinematics have been calculated using the theoretically obtained densities 
for some of the nuclei and compared with available experimental data. 
The total cross cross sections for elastic scattering have also 
been calculated for all the nuclei studied showing a large increase
for the halo nucleus $^{14}$Be. The nuclei have also been 
investigated for deformation. The nuclei $^{10}$Be and $^{16,18,20}$C
are observed to be deformed in their ground state.

\end{abstract}
\pacs{21.60.Jz, 21.10.Gv, 17.20.+n, 27.30.+t}

\maketitle
\clearpage
\section{Introduction}
The last fifteen years have been an exciting time for nuclear physics.
With the giant leaps in detection systems and accelerator technologies,
particularly with the availability of radioactive ion beams, the old
theories have been severely tested as never before. The limits of nuclear
stability are now being probed and yielding surprising results. 
Major surprises in low energy nuclear structure include the 
disappearance of the normal shell closures 
observed near the stability valley along with the
emergence of new magic numbers
and neutron halo in nuclei very close to the 
drip line\cite{exotic}. The effect of the halo may be observed in different 
reactions involving these nuclei. In particular, though electron scattering 
is the most direct probe of nuclear density, 
it is difficult to apply in nuclei far away from the valley of
stability. Elastic proton scattering in inverse kinematics
provides a test for the calculated densities\cite{EP}. 
In the present work, we study the structure of exotic even-even Be and 
C nuclei and calculate the elastic proton scattering cross sections 
using the theoretical densities.

Beryllium and carbon isotopes show a number of interesting features. 
The neutron drip line in beryllium is at 
$^{14}$Be which is known to be a two-neutron halo nucleus. 
The drip line nucleus for carbon isotopes is $^{22}$C which 
 may also have a similar halo.
It has also been suggested that N=16 is a new spherical magic number
\cite{PRL}. 
Our aim is to see how well Relativistic Mean Field (RMF) calculations 
can explain these different observed features in these nuclei.

There has been a number of nonrelativistic mean field calculations for the
binding energy and radius in these nuclei\cite{Sag}.
Patra\cite{Pat} has studied a number of light nuclei including Be and C 
isotopes using RMF approach. Ren \etal have studied Be nuclei
using density dependent RMF \cite{Ren1} and C \cite{Ren2} nuclei with RMF.
The radius and binding energy of $^{14}$Be have been reproduced in their
calculation. Gmuca have studied various Be isotopes using the relativistic
mean field approach\cite{Gmuca}.
Sharma \etal\cite{Sh} have studied the exotic carbon isotopes in
a relativistic theory.
Sugahara \etal \cite{Su} have also studied these nuclei using 
relativistic and non-relativistic theories. 
Recently, deformed relativistic Hartree Bogoliubov (RHB) calculation employing 
the force NL3\cite{NL3} 
in oscillator basis has also been 
performed to describe $^{11-14}$Be and $^{14-22}$C nuclei\cite{Lal} among 
others.
Most of the available calculations in this region use
the harmonic oscillator basis. 
Spherical RHB approach has also been used \cite{Po} to study C nuclei.

Though there has been a number of calculations in the RMF approximation for
Be and C nuclei, most of them use the harmonic oscillator basis. However,
the basis expansion method may not be able to describe loosely bound states 
in halo nuclei. Calculations in RHB approach in r-space are also available.
However, they are very involved and time consuming. We
have used the co-ordinate space RMF+BCS approach to study these nuclei
and to compare the results with thaose of RHB calculation
in co-ordinate space.

\section{Theory}

RMF\cite{RMF} calculation is now a standard tool to investigate the 
structure of the nucleus. It has been able to explain different features 
of stable and exotic nuclei like ground state binding energy, deformation, 
radius, excited states, spin-orbit splitting, neutron halo, etc\cite{RMF1}. 
Relativistic calculations have been known to give good description of 
nuclei near the drip line. For example, it has been possible to describe the 
halo in even the very light nuclei $^{11}$Li\cite{Meng}. This could be done 
without any artificial adjustment of the potential as required in the previous 
nonrelativistic calculations. Our aim is to see whether relativistic mean field
calculations can also correctly describe the different features in Be and C 
nuclei. It is worthwhile 
to note that in $^{14}$Be, the binding energy and radius could be reproduced
in an RMF calculation\cite{Ren1}.
The starting point is the relativistic Lagrangian for point nucleons 
interacting via exchange of the
scalar-isoscalar meson $\sigma$, the vector-isoscalar meson $\omega$, the
vector-isovector meson $\rho$ and the photon. RMF is known to give a good
description of spin orbit splitting and is thus ideally suited for
investigating the magic numbers in nuclei away from the stability valley. 
 In recent years, efforts have been made to develop a energy
density functional which will be applicable to all nuclei in their ground as 
well as excited states and to  nuclear matter. Within the relativistic
framework, effective interactions have been constructed
with density dependent meson nucleon couplings\cite{dd} for this purpose.
These recent developments are motivated by the fact that the success of the 
RMF approach is now  explained from the point of view
of effective field theory and the density functional theory. For example,
the nonlinear terms in the Lagrangian are now considered to introduce
additional density dependence in the energy functional. The parameters
of the Lagrangian have been obtained by fitting different experimental
observations and may be interpreted in this approach to already contain the
vacuum contributions.
 In
quantum hadrodynamics effective field theoretical Lagrangians explicitly
include the basic symmetries of QCD and thus may be considered as
its true representation in the low energy nuclear physics.
The readers are referred to recent literature
\cite{Serot} for additional details.
         
In the conventional RMF+BCS approach for even-even nuclei, the Euler-Lagrange 
equations obtained are solved under the assumptions of classical meson
fields, time reversal symmetry, no-sea contribution, etc. Pairing is introduced
under the BCS approximation. Both constant gap and constant strength methods
as well as other approaches in pairing have been used in different works. Very
often the resulting equations are solved\cite{Gam} in a harmonic oscillator 
basis. However, in exotic nuclei, the basis expansion method using 
harmonic oscillator, because of its incorrect asymptotic properties, faces
problems in describing the loosely bound halo states. A solution of the 
Dirac and Klein Gordon 
equations in co-ordinate space may be preferable to describe the weakly bound 
states. Because these nuclei studied are very close to the drip line, one has 
to consider the effect of the positive energy states also. In this work, we 
have calculated the resonant state by studying localization of the scattering 
wave function except for the $\nu s_{1/2}$ state. This method has been 
applied in the nonrelativistic Hartree Fock \cite{conpair} as well as 
relativistic mean field formalism\cite{con}.
 The $\nu s_{1/2}$ state could 
not be localized because there is no Coulomb or centrifugal barrier for this 
state. Thus we have to use the box normalization condition for the positive 
energy $\nu s_{1/2}$ state which occurs only in $^{10}$Be among the nuclei 
studied in the present work. 
These positive energy levels are of finite width whose effect in pairing can be 
incorporated following Ref. \cite{conpair}. However, because the contribution 
of these levels are expected to be small, we have assumed these levels to be 
of zero width at the resonance energy.
In the very light mass region, where we are interested, pairing energy is very 
small.  We have followed two procedures in pairing, both in 
constant gap approximation. In one 
we have taken the pairing gaps as $\Delta_p=\Delta_n=0.2$. This prescription 
has been followed by Ren \etal\cite{Ren2}. In the other
we have adjusted the gap parameters so as to reproduce 
the pairing energy obtained in the spherical RHB calculation described below.
 This method has been successfully followed in many works\cite{Bh}.
We call these two procedures RMF-I and RMF-II, respectively.  We find 
that there is very little difference between the  two approaches.

A more accurate treatment of the drip line nuclei involves RHB approximation
which has been studied in a number of previous works. We have compared the 
results of our RMF+BCS calculation with that of RHB approach  
using the same force NLSH. The RMF+BCS
calculation is much simpler and less time consuming compared to a full RHB 
calculation. We want to compare the results of the two approaches, particularly
for the density distribution. 
For the RHB calculation, we have used the code spnRHBfem\cite{cpc}.
For the finite range interaction, the $J=0$ part of 
nonrelativistic Gogny interaction D1S\cite{D1S} has been chosen. 

The RMF+BCS approach in co-ordinate space was modified to study deformed nuclei also\cite{GG}. We have adopted this method to study the deformation of these 
nuclei.   The quadrupole deformation parameter $\beta_2$ is calculated from the
total quadrupole moment using the relation

\bea Q_{n,p}= \sqrt{\frac{16\pi}{5}}\frac{3}{4\pi}(N,Z)R_0^2\beta_{2n,p}\eea
with $R_0$=1.2A$^{1/3}$fm.

We have applied the force NLSH\cite{NLSH} in the co-ordinate space RMF approach 
to calculate the ground state properties in neutron rich Be and C nuclei. 
 We have also checked our results with the force NL3. However, 
as discussed later, we find that
the agreement in binding energy, particularly for the drip line nuclei,
is better in the case of the NLSH force.

Although electron scattering is the most direct method for measuring the 
density in stable nuclei\cite{ES}, it is difficult to apply in regions far 
away from the valley of stability. Elastic proton scattering in inverse 
kinematics, alternatively, also 
provides a test for the calculated densities\cite{EP}. We have calculated 
the elastic 
scattering cross section for scattering of the nuclei from proton target at 
55$A$ MeV energy with the optical model potential (OMP) generated in a 
semi-microscopic approach. The OMP is obtained using the effective interaction 
derived from the nuclear matter calculation of Jeukenne, Lejeune, and Mahaux 
(JLM)\cite{JLM74} in 
the local density approximation (LDA) by substituting the nuclear matter density
with the calculated density distribution of the finite nucleus. Further 
improvement is incorporated in terms of the finite range of the effective
interaction by including a Gaussian form factor. The calculation has been 
performed with the computer codes MOMCS\cite{MOMCS} and ECIS95\cite{ECIS}
assuming spherical symmetry. We have used the global parameters for the 
effective interaction and the respective default normalizations for the 
potential components from Refs. \cite{MOMCS} and \cite{MOM} with Gaussian 
range values of $t_{real}=t_{imag}=1.2$ fm. No search has been performed on 
any of these parameters. It has been shown previously that JLM calculation 
can reproduce the proton plus unstable nucleus elastic scattering, when a 
realistic nuclear matter density distribution is used \cite{EP}. 

\section{Results}

We have studied the structure of $^{10,12,14}$Be and $^{14,16,18,20,22}$C.
In Table \ref{tab1} our results for binding energy and radius values 
in the spherical limit are given and compared with experimental measurements 
wherever available. All the theoretical values in this table have been 
calculated using the force NLSH.
The calculated results for binding energy are in reasonable 
agreement with experimental measurements. Later we will show that this agreement
improves with the inclusion of deformation degree of freedom. The results for 
different type of 
radii are in excellent agreement with experimental values in most of the cases. 
One can also see that the results of RMF-I calculation  do not significantly 
differ from that of RHB calculation except for the radius of $^{14}$Be where 
the latter is closer to the experimental value.  
Sandulescu \etal\cite{con} have pointed out that this difference is
generally common near the drip line and can be attributed to the different ways
of pairing calculation in the two methods. The occupancies of narrow resonances
with high angular momenta is higher in RHB calculation. This is a consequence of
the large energy cut off employed in the RHB (or HFB) approach which makes the
Fermi sea more diffuse, thus increasing the scattering to loosely bound narrow
resonances with high angular momenta. The RMF calculations, on the other hand,
predict higher occupancy of broader low angular momentum resonances. The radius
near the drip line is very sensitively dependent on the occupancy of the
localized orbits. The high spin states are more localized due to
larger centrifugal barrier. Increased occupancy for them translates into
smaller radius for RHB calculation.

As expected, the RMF-II calculation gives a better agreement with the RHB 
calculation. In all the other features studied in the present approach in the 
spherical limit the three methods agree very well among themselves and we 
present the results of RMF-I only for them unless otherwise mentioned.

The single particle neutron levels in Be are given in Fig. 1. A level 
inversion occurs with the $2s_{1/2}$ state coming down below the $1d_{5/2}$ 
state. The former becomes weakly bound in $^{12,14}$Be. This inversion is 
essential for the nucleus $^{14}$Be to be bound. However, neither the  
RMF+BCS scheme, nor the RHB approach can predict the 
parity inversion in neutron rich Be nuclei, which has been observed in 
$^{11}$Be. This inversion also could not be explained by a full scale 
shell model calculation\cite{SM} and the authors of that work have suggested 
that the effect of three body forces should be included to explain the 
phenomenon.  Although the RMF forces contain contributions from higher
body forces, both the presently used parameterizations, {\em i.e.} NLSH and 
NL3 fail in this regard.

In Fig. 2, we plot the nucleon densities in 
$^{12,14}$Be for a  spherical calculation in co-ordinate space as well as
the RHB result. The RMF+BCS densities are indicated by the solid lines, 
and the RHB  densities, by the dashed lines. The proton densities are 
very similar in both the nuclei. The neutron halo in 
$^{14}$Be is clearly seen in both the calculations. 

The calculated single particle levels in $^{18,20,22}$C are shown in Fig. 3.
The RMF-I results  are compared with those  of RHB calculation. One can see 
that for the negative energy levels, the RMF+BCS calculation agrees very well with the
more involved RHB approach.
A difference 
between the single particle neutron structures of Be and C isotopes is 
readily seen. 
In C isotopes, the level 
inversion between the $1d_{5/2}$ and the $2s_{1/2}$  single neutron levels 
does not
occur though the latter  comes very close to the former. Another
important difference is the binding energy of the last filled level. The
$2s_{1/2}$ state is bound by more than 3 MeV. Hence, the wave function of 
this state does not extend to a very large value unlike the results
of calculation in $^{14}$Be and the predicted neutron radius is actually
smaller than that expected for a halo nucleus. 

The nucleon densities in $^{16,18,20,22}$C have been shown in Fig. 4. 
Once again, the results of the RMF+BCS and the RHB calculations 
agree very well. Both the calculations 
indicate that the neutron density distribution
of $^{20}$C and $^{22}$C are not substantially different. This is
another aspect of the fact that according to our calculation, the nucleus 
$^{22}$C does not have a two neutron halo. It has been suggested 
that the neutron number N=16 is the new magic number in neutron rich nuclei.
In the present work, the gap between the $2s_{1/2}$ and the $1d_{3/2}$ level 
comes out to be nearly 5 MeV in accordance with the above prediction.

 To check whether the results in the present calculation depend on the
particular force chosen, we have compared the results of the NLSH force with 
those of another similar nonlinear force, NL3.
We have followed the procedure of RMF-I, {\em i.e.} 
performed a constant gap calculation with $\Delta_n=\Delta_p=0.2$.
As mentioned earlier, the agreement in the case of binding energy 
obtained with the NL3 force is poorer, particularly as the neutron
number increases. Thus, in the case of $^{12}$Be, the predictions
from NLSH and NL3 forces are 4.992 MeV and 5.086 MeV, respectively.
Similarly, for $^{22}$C, the corresponding values are 5.568 MeV and
5.662 MeV, respectively. However, the general pattern of the ground state
properties, including the level inversion mentioned above, and the density are
the same for both the forces.

In Fig. 5, the results of the model calculation for angular distribution of 
elastic scattering of $^{12,14}$Be and $^{20,22}$C from proton target in 
inverse kinematics with density distributions taken from the RMF calculation 
have been plotted. 
We have come across only one experimental result for elastic scattering
among all the nuclei, {\em viz.} for $^{12}$Be at an energy of 55A MeV\cite{EL}.
For all the results above the energy is taken to be 55A MeV.
The theoretical results are compared with the 
experimental data taken from \cite{EL}. As is apparent from the above 
discussion, the density patterns obtained from RMF and RHB calculation are 
very similar. Hence, we find that the scattering cross sections obtained
using those density profiles are nearly identical in the two cases and 
have shown the scattering cross section in the
RMF approach only.  One can see that the trend of the 
scattering angular distribution can be satisfactorily explained by the present 
calculation without any further adjustment of the parameters of the effective 
interaction. No other experimental
data is available for elastic proton scattering for the nuclei studied in the 
present work.  In Fig. 6, we have 
plotted the total cross sections for elastic proton scattering in inverse 
kinematics. The beam energy in each case is $55A$ MeV. The smooth lines
show the $A^{2/3}$ behaviour. Although in both the chains, the calculated 
values show an increase over the $A^{2/3}$ behaviour, one can see that for 
$^{14}$Be, there is a large increase in the total cross section over the 
corresponding value for $^{12}$Be. In the case of C nuclei, the rise is more 
gradual, even for $^{20}$C-$^{22}$C. This smooth rise in the cross section 
is due to the fact that our calculation does not predict a two neutron halo in 
$^{22}$C. Thus an experimental measurement of total elastic scattering 
cross section can verify the presence or absence of two neutron halo in 
the dripline Be and C isotopes. 

We have also studied the nuclei for deformation in RMF-I and RMF-II formalism.
The nuclei $^{12}$Be and $^{14,22}$C are found to be spherical
in agreement with the fact that the neutron number N=8 and N=16 are 
magic numbers.  In this regard, our calculation agrees with the RHB 
results of Lalazissis \etal\cite{Lal}. We also have observed $^{14}$Be to 
be spherical.
All the other nuclei show varying degree of deformation.
The results of our calculation for binding energy and quadrupole deformation 
after the inclusion of deformation are 
presented in Table 2.  For deformed nuclei, the calculation selfconsistently 
converges to 
the two coexistent minima, prolate and oblate, according as one starts
 with a positive or a negative initial deformation.
One can see that in all these nuclei, the agreement 
between the calculated binding energy and experimental measurements
improve for deformed solutions. Also, in most cases the proton and neutron
deformation are substantially different from each other.
In all the deformed nuclei studied, solutions for prolate
and oblate deformation are very close in energy. 
We find that our results agree with that of the deformed RHB calculation
of \cite{Lal} except in a few cases as discussed. 
The nucleus $^{10}$Be,  which has not been studied in Ref. \cite{Lal}, comes 
out 
to be strongly deformed. Here, the proton deformation is much larger than 
the corresponding neutron one. 
In $^{14}$Be, because
of the level inversion, the last two neutrons occupy the $2s_{1/2}$ level
instead of the deformation driving $\Omega$=5/2 orbital of the $1d_{5/2}$
level as expected from level ordering observed near stability valley. 
Hence its ground state comes out to be spherical in contrast to Ref. \cite{Lal},
where the ground state is obtained as strongly deformed. In $^{16}$C, the
prolate and the oblate solutions come out to be nearly degenerate.
This was observed in Ref. \cite{Lal} also. Similarly, in $^{18}$C, although the 
ground state comes out to be prolate, the binding energy of the oblate solution 
is only about 150 keV less. The nucleus $^{20}$C is again observed to be oblate.
 In contrast, the deformed RHB calculation \cite{Lal} suggests that both
$^{18,20}$C are oblate in their ground states. Lalazissis \etal have
noted that because of the close lying self-consistent minima, it is not always
possible for the mean field theories to accurately predict the sign of the 
deformation.
In all the deformed C isotopes, proton distribution is very weakly deformed while 
the neutron distribution, except for the case of the prolate solution in 
$^{20}$C, show moderate to large deformation.  For comparison with
the density obtained in the spherical solution, we plot in Fig. 7, the 
monopole (L=0) and the quadrupole (L=2) components of the neutron and proton 
densities in $^{16}$C from the deformed calculation as well as the densities 
obtained in the spherical approach, both using RMF-I.
One can see that the monopole components of the deformed distribution is 
similar to the spherical results 
except at the core where the latter is slightly depressed for neutrons. At 
larger distances the neutron distribution has a substantial contribution coming from the quadrupole component.

 To check whether the disagreement in the ground state shape in
$^{14}$Be between the present calculation and \cite{Lal} is due to 
the different force used or the essentially different methods adopted,
we have employed the NL3 force in our calculation. In the resultant 
prolate solution, the proton distribution is nearly spherical. On the other 
hand, the neutron distribution shows a prolate deformation with 
$\beta_{2n}=0.16$
which corresponds to a small mass deformation $\beta_2=0.11$. In 
comparison, the RHB calculation of Lalazissis \etal \cite{Lal}, have
predicted a very large mass deformation (nearly 0.4). One of the reasons
of this large difference may be the fact that the RHB calculation, as
mentioned earlier, involves a large number of positive energy levels
while the RMF+BCS calculation includes only a few levels around the 
Fermi energy.

Overall, we find that our results for binding energy and radii in all 
the C isotopes studied in the present calculation are in better agreement 
with experimental or empirical values than the oscillator basis calculation\cite{Sh} which uses the forces NLSH or TM1. In $^{22}$C, because of the new magic 
number N=16, our calculations predict a spherical ground state. It is also 
better than or comparable to other relativistic calculations\cite{Pat,Ren2,Lal} 
in oscillator basis in the Be isotopes using various forces in this regard.
For example, the deformation values obtained in Carbon isotopes in the present 
approach are more in agreement with the RHB results than the oscillator basis
calculations.
Thus we may conclude that near the drip line co-ordinate space calculation in 
many cases is better than basis expansion approach and is comparable to the 
more involved RHB approach.

\section{Summary and Conclusions}

The structure of neutron-rich C and Be nuclei have been studied in 
co-ordinate space RMF calculation and compared with results of RHB 
approach. The position of the neutron drip 
line is correctly predicted in both the elements. In Be isotopes, the 
$\nu 2s_{1/2}$ level comes below the $\nu 1d_{5/2}$ level, providing the
drip line nucleus $^{14}$Be with a two neutron halo. On the other hand, 
in $^{22}$C, this inversion does not occur. Moreover, in $^{22}$C the 
last filled level
is bound by about more than 3 MeV and there is no two-neutron halo. 
The much simpler RMF+BCS approach agrees very well with the RHB results.
The
densities calculated have been used to construct optical model 
potentials for proton scattering. The calculated differential cross section 
for elastic proton scattering in inverse kinematics compares favourably
with experiment. The total elastic scattering cross section values for 
different nuclei have also been calculated. It
shows a sudden increase at the neutron halo nucleus $^{14}$Be.
We have also studied the effect of including the deformation degree of freedom.
The nuclei $^{10}$Be and $^{16,18,20}$C come out to be deformed. The agreement
of the prediction for ground state binding energy with experimental 
measurement improves substantially with the inclusion of deformation.
 Here again, the agreement with
deformed RHB calculation is noteworthy. In many cases, the co-ordinate space
calculations are seen to be better than basis expansion approach.

\noindent{\bf Acknowledgment}

A part of the calculation was done using the computer facilities provided 
under the DSA Programme by the University Grants Commission, New Delhi.

\clearpage
\begin{table}
\caption{Binding energy and radius in Be and C isotopes in the spherical 
approach for NLSH force. Experimental binding energy values are from the compilation 
\cite{AWT}. Experimental r.m.s. radii values are from \cite{OST} and are 
results of Glauber model analysis in the optical limit. Experimental neutron 
radii for C isotopes are from \cite{rn}.\label{tab1}}
\center
\begin{tabular}{ccclll}\cline{1-6}
$^AZ$& &B.E./A(MeV)&\multicolumn{3}{c}{Radius(fm)}\\
&&&r$_p$&r$_n$&r$_{rms}$\\\hline
$^{10}$Be &Expt.& 6.498 && & 2.30(2)\\
          &RMF-I& 6.192 &2.19 & 2.42 & 2.33\\
          &RMF-II& 6.192 &2.19 & 2.42 & 2.33\\
          &RHB& 6.188 & 2.19&2.42&2.33\\
$^{12}$Be &Expt.& 5.721& &&2.59(6)\\ 
          &RMF-I& 5.855&2.26 & 2.73 & 2.58 \\ 
          &RMF-II& 5.847&2.26 & 2.72 & 2.58 \\ 
          &RHB&5.845&2.26&2.72&2.58\\
$^{14}$Be &Expt.& 4.994 && &3.16(38)\\
          &RMF-I& 4.992&2.27&4.04 & 3.62\\
          &RMF-II& 4.985&2.27&4.05 & 3.63\\
          &RHB& 4.955 &2.27&3.69&3.35\\
$^{14}$C &Expt.&  7.520&&2.70(10)&2.30(7)\\ 
         &RMF-I& 7.616& 2.37 & 2.56 & 2.48\\ 
         &RMF-II& 7.616& 2.38 & 2.56 & 2.48\\ 
         &RHB&7.612&2.38&2.56&2.48\\
$^{16}$C &Expt.&  6.922&&2.89(9)&2.70(3)\\ 
         &RMF-I& 6.780& 2.39 & 2.93 & 2.74\\ 
         &RMF-II& 6.765& 2.39 & 2.88& 2.71\\
         &RHB&6.765&2.39&2.85&2.69\\
$^{18}$C &Expt.&  6.426&& 3.06(29)&2.82(4)\\ 
         &RMF-I& 6.220&2.41 & 3.04 & 2.84\\ 
         &RMF-II& 6.211&2.41 & 3.02 & 2.82\\ 
         &RHB&6.206&2.41&3.01&2.82 \\
$^{20}$C &Expt.&  5.959 && &2.98(5)\\
         &RMF-I& 5.843&2.43 & 3.16 & 2.96\\ 
         &RMF-II& 5.844&2.43 & 3.12 & 2.93\\ 
         &RHB&5.843&2.43&3.12&2.93\\
$^{22}$C &Expt.&  5.440$^1$\\
         &RMF-I& 5.568& 2.45 & 3.38 & 3.15\\ 
         &RMF-II& 5.568& 2.45 & 3.38 & 3.15\\ 
         &RHB&5.565&2.45&3.36&3.14\\\hline
\end{tabular}

$^1$ Estimated value
\end{table}

\begin{table}
\caption{Calculated binding energy and deformation($\beta)$ in Be and C 
isotopes. NLSH force has been used.
\label{tab12}}
\center
\begin{tabular}{crrrrrrrr}\hline
Nucleus & \multicolumn{4}{c}{RMF-I}&\multicolumn{4}{c}{RMF-II}\\
&$\beta_{2p}$&$\beta_{2n}$&$\beta_2$& B.E./A 
&$\beta_{2p}$&$\beta_{2p}$& $\beta_2$& B.E./A \\
&&&& MeV &&&  & MeV \\\hline
$^{10}$Be & 0.36 &0.14 & 0.23 &6.398& 0.37 & 0.14&0.23&6.415\\
          &-0.24&-0.16&-0.19&  6.365 & -0.25 &-0.17 & -0.20&6.372\\
$^{16}$C &0.08  & 0.46  &0.32  & 6.888 & -0.11&-0.24 &-0.19&6.904\\
         &-0.11 & -0.24 &-0.19 & 6.879 & 0.08 & 0.47 &0.32 &6.902\\
$^{18}$C &0.09  & 0.40  & 0.30 & 6.387 & 0.09 &  0.39&  0.29&6.375\\
         &-0.12 & -0.39 & -0.30& 6.381 & -0.11& -0.37&-0.28 &6.365\\
$^{20}$C &-0.11 & -0.30 & -0.25& 5.968 & -0.12 & -0.31 & -0.25 & 5.983\\
        &  0.04 & 0.14 & 0.11 &  5.865 &  0.03 & 0.12 &0.09&5.865\\
\hline
\end{tabular}
\end{table}
\clearpage

\centerline{\bf List of Figure captions}

\vskip 1cm

\parindent 0.0cm
Fig. 1 : Calculated single particle neutron states in  
$^{10,12,14}$Be in the spherical approximation. See text for details.

Fig. 2 : Calculated proton and neutron densities in $^{12,14}$Be in the 
spherical approximation. Neutron and proton densities are indicated by N and P,
respectively. The solid (dashed) line represents results of RMF+BCS(RHB)
calculations.

Fig. 3 : Calculated single particle neutron states in 
 $^{18,20,22}$C in the 
spherical approximation.

Fig. 4 : Calculated proton and neutron densities in $^{16,18,20,22}$C in the 
spherical approximation. See caption of Fig. 2 for details.

Fig 5. : Partial cross section for the elastic proton scattering 
in inverse kinematics. Energy of the projectile is 55A MeV.
Theoretical results are connected by the solid line. Experimental values 
are from \cite{EL}.

Fig 6 : Total cross section for elastic proton scattering of different 
C and Be nuclei in inverse kinematics studied in the present work.
 
Fig. 7 : Neutron and proton densities obtained in deformed and spherical
calculation in $^{16}$C. Neutron and proton densities are indicated by
N and P, respectively. See text for details.
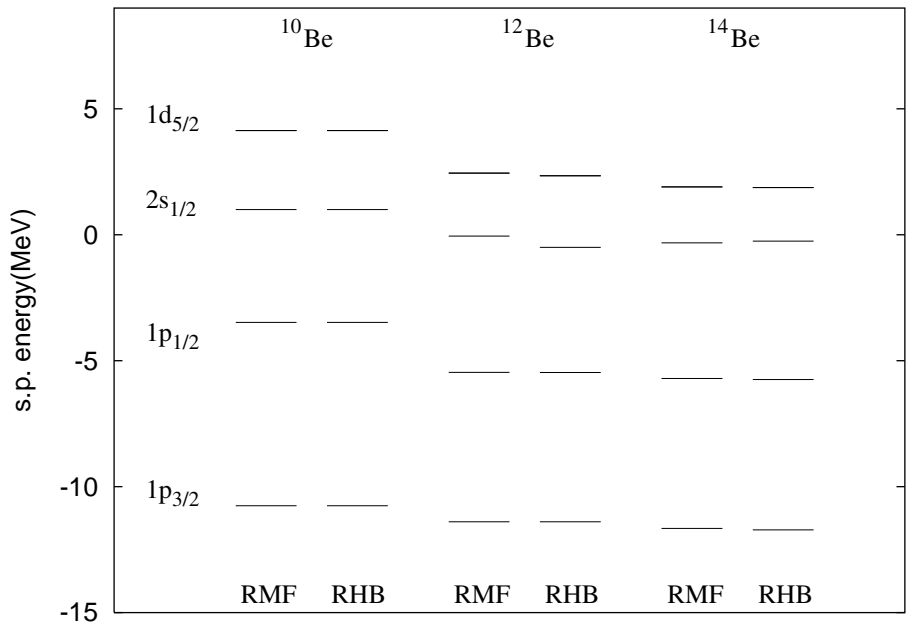
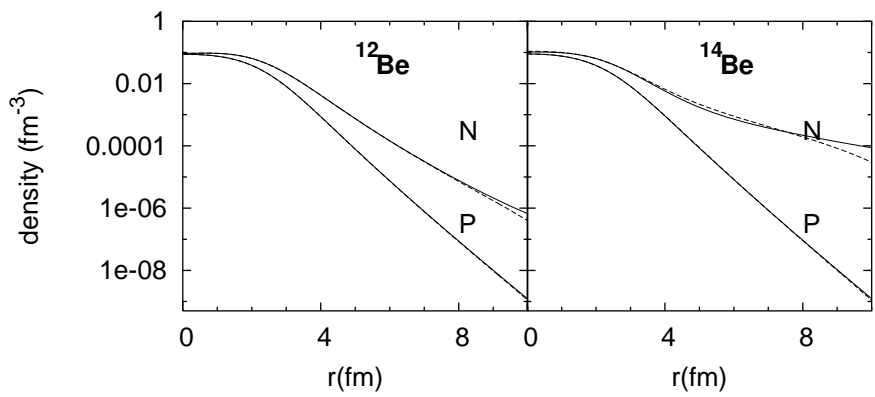
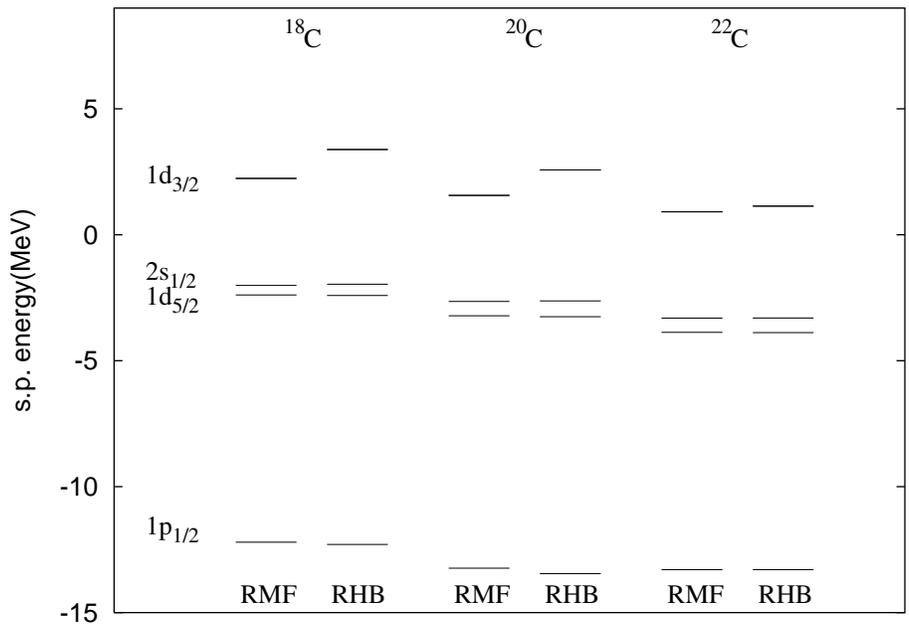
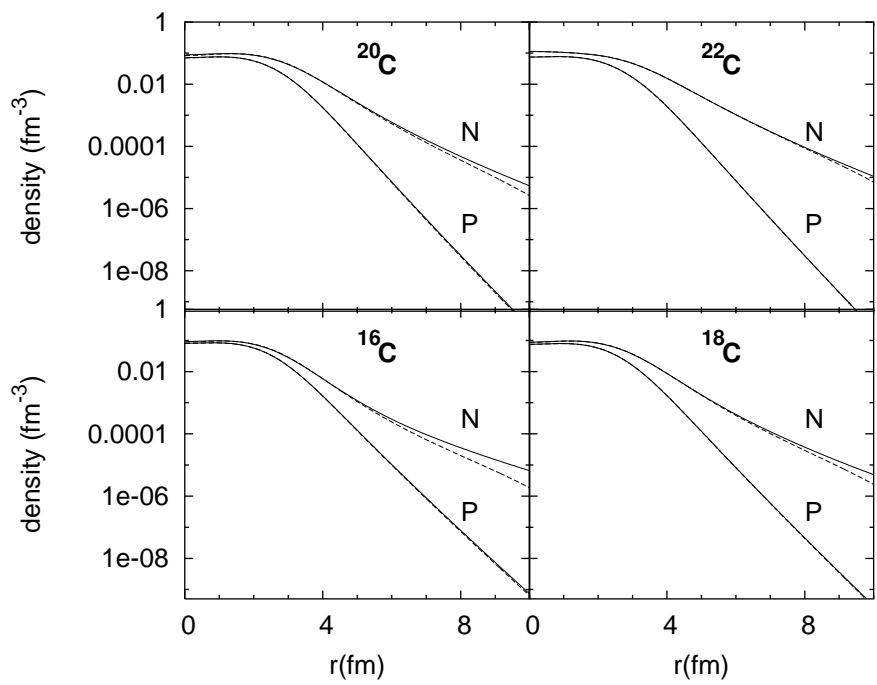
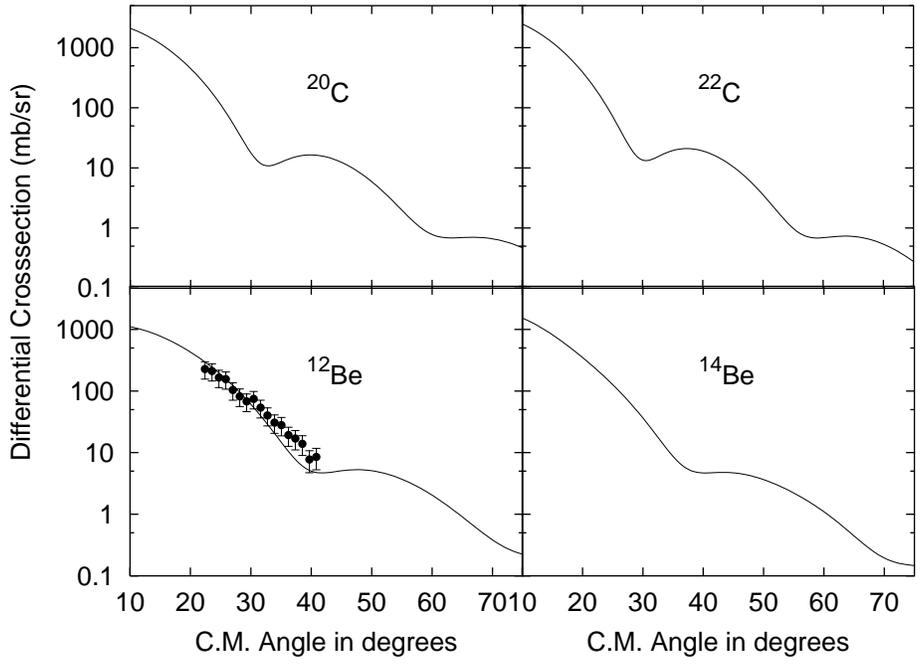
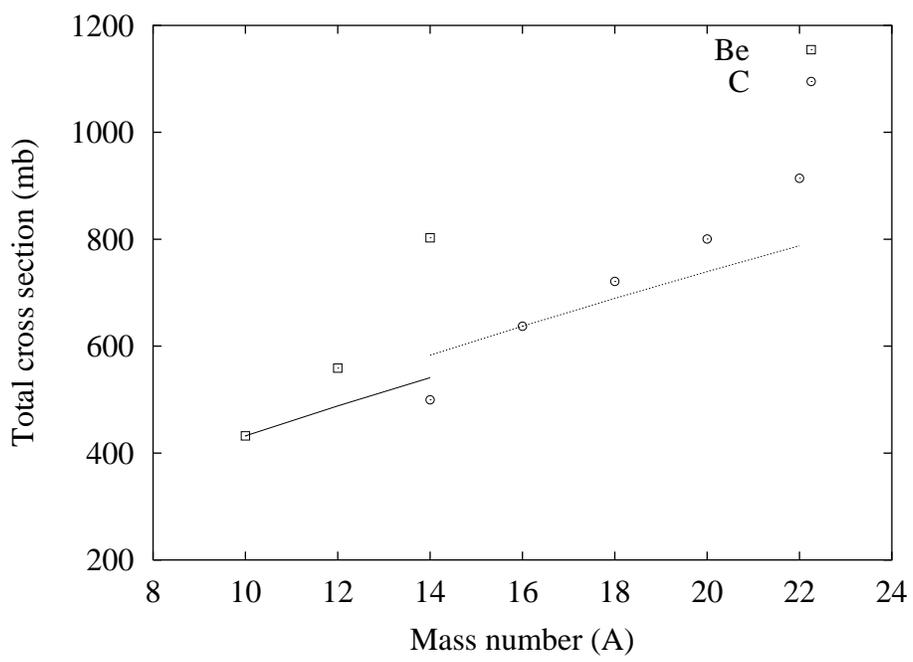
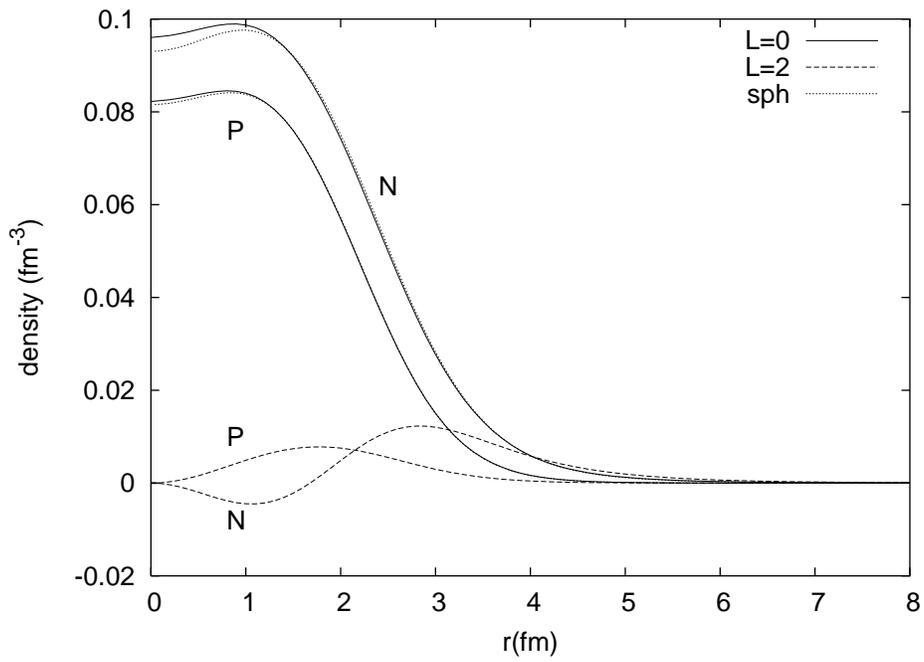
\end{document}